\documentstyle[multicol,graphics,aps]{revtex}

\begin{document}

\title{Probing the potential landscape inside a two-dimensional electron-gas}
\author{J.J.\ Koonen\cite{josaddr}, H.\ Buhmann, and L.W. Molenkamp} 
\address{Physikalisches Institut, Universit\"at W\"urzburg, Am Hubland, 97074 W\"urzburg, Germany} 

\maketitle

\begin{abstract} 

We report direct observations of the scattering potentials in a two-dimensional electron-gas using 
electron-beam diffaction-experiments. The diffracting objects are local density-fluctuations caused by 
the spatial and charge-state distribution of the donors in the GaAs-(Al,Ga)As heterostructures. The 
scatterers can be manipulated externally by sample illumination, or by cooling the sample down under 
depleted conditions.\\


\end{abstract}

\begin{multicols}{2}

The high electron mobilities that can be obtained in the two-dimensional electron-gas (2DEG) in GaAs-
Al$_{x}$Ga$_{1-x}$As heterostructures continue to fascinate the community \cite{saku96,umansky97}. At low 
temperatures the mobility of electrons in a perfect 2DEG is, in principle, limited by ionized donor 
scattering. The donors are located in the doping layer, some tens of nanometers away from the 2DEG. Due 
to the random distribution of donor atoms, the scattering potential is not homogeneous \cite{nixon}. 
However, there are theoretical \cite{efros} and experimental \cite{coleridge91,buks} indications that 
spatial correlations between donors in different charge states reduce the ionized donor scattering and 
thus enhance the mobility.

These different charge states exist because for structures with a certain content of Al ($x\ge 0.2$) the 
electronic ground-state of the Si-donor is two-fold  \cite{queisser,chadi,bednarek}: First, a shallow 
donor state, which is associated with a normal substitutional lattice-site and a binding energy of 
approximately 7 meV (d$^0\Rightarrow\, $d$^+ + e$). Second, a more localized, deep donor level with a 
binding energy of $\approx$ 160 meV, the DX-center, which derives from lattice distortions at or near the 
donor site. In fact, the latter is a negatively charged donor state, DX$^-$, which in contrast to the 
neutral and positively charged DX-states is stable with respect to the equivalent shallow donor state. At 
low temperatures ($T < 130$ K) DX$^-$-states become stable against thermal dissociation (DX$^- 
\Leftrightarrow\, $d$^0 + e$).

Several aspects of the high mobilities of a 2DEG can now be explained by invoking spatial correlations 
between donors in different states, d$^+$ and DX$^-$\cite{coleridge91,buks}, where the roughening of the 
potential caused by donors in one state is effectively screened by donors in the other state. The 
electrostatic interaction should lead to regions of several tens of nanometer in diameter where all the 
donors are in one state \cite{efros}. These correlations lead to regions of reduced density in the 2DEG 
below the donors\cite{nixon,efros}. The correlations can be altered externally by sample 
illumination\cite{chadi,fletcher,hayne}, causing a dissociation of DX$^-$-centers ($E_{\rm excite} \ge 
1.2$ eV) and "bias-cooling", i.e.\ cooling the sample down while the 2DEG is depleted by an applied gate 
voltage\cite{buks,long,skuras,coleridge} (this is an alternative to prevent the DX$^-$-formation). In the 
experimental studies performed sofar\cite{coleridge91,buks}, the mobility of a 2DEG was inferred from 
standard bulk conductivity measurements. Such experiments probe an averaged scattering potential and 
therefore do not yield experimental information about the local distribution of shallow donors and DX-
centers. Moreover, the evidence for the occurence of donor correlations is only indirect.

In this article we use a collimated electron beam, injected and detected via quantum 
point-contacts (QPC), as a local probe for the scattering potentials in a 2DEG, which, as we will 
demonstrate, are the regions of reduced density caused by the donor state correlations. The observed 
interference patterns are analysed using a theoretical model based on a technique developed by M.\ Saito 
et al.\cite{saito}, extended to the situation where impurities are present in the 2DEG region. This model 
allows for a deduction of the size and location of the scattering potentials. Experimentally, the donor 
configurations are changed by illumination and bias-cooling techniques.

For the experimental investigations, several gate-defined nanostructures in conventional 
GaAs-Al$_{0.33}$Ga$_{0.67}$As-heterojunction are used. The relevant part of the layer structure consists 
of 400 nm undoped GaAs, 20 nm undoped Al$_{0.33}$Ga$_{0.67}$As (spacer-layer), 38 nm $1.33\times 10^{18}$ 
cm$^{-3}$ Si-doped Al$_{0.33}$Ga$_{0.67}$As, and 17 nm undoped GaAs (cap layer). Typical values for the 
carrier density and moblilty are $n=1.5\dots 2.3\times10^{15}$ m$^{-2}$ and $\mu=60\dots 150$ m$^2$ 
(Vs)$^{-1}$. A schematic topview of a typical gate structure is given in Fig.~1a. Schottky-gates form two 
opposite QPCs (injector and detector), separated by a distance of typically $L=4$ $\mu$m. In some samples 
the area in between the QPCs is partly covered by an additional Schottky-gate (light grey regions, 
Fig.~1a). The conductance of the QPC can be adjusted in such a way that only $N$ conducting modes are 
transmitted ($N = G\, h / 2e^2$). A small, low frequency ac-voltage ($V_{\rm ex}\approx$ 100 $\mu$V, 13 
Hz) is applied to the ohmic contact I$_{\rm i}$, injecting an electron-beam into the 2DEG. By using lock-
in techniques, the voltage drop over the detector QPC is measured (contacts: V$_{\rm c}^1$ and V$_{\rm 
c}^2$). Due to the smooth boundaries of an electrostatically defined QPC, the injected electrom beam is 
collimated\cite{molenkamp}. In a weak magnetic field perpendicular to the 2DEG plane, the electron beam 
is deflected by the Lorentz force. 
\begin{figure}
\begin{center}
\resizebox{6.5cm}{10cm}{\includegraphics{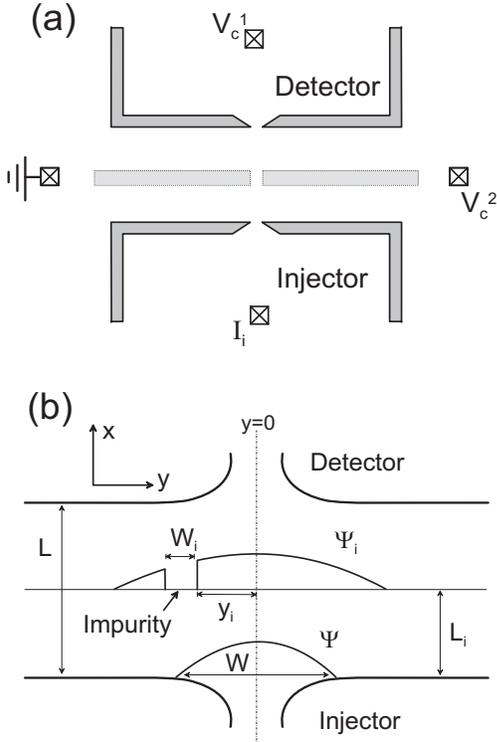}}
\begin{minipage}{8cm}
\caption{Schematic view of the sample: (a) shows the gate structure and ohmic contacts (crossed squares) 
of the device. Dark  grey: QPC gate; light grey: optional gate used for the bias-cooling experiment. (b) 
defines lengths and the coordinate system and displays the  electron beam wavefunction $\Psi$ at the exit 
of the injector quantum pointcontact (QPC) and after scattering, $\Psi_i$.}
\label{sample}
\end{minipage}
\end{center}
\end{figure}
Therefore, the measured non-local resistance, $V_c/I_i$, as a function of magnetic field provides 
information about the beam profile. Experimentally, the measured beam profile is not smooth, but rather 
exhibits additional structure (see Fig.~2, 3, 4, and Fig.~1 Ref.~\cite{molenkamp}). Also other 
groups\cite{okada,roukes} presented data exhibiting these features, but their origin has not been 
discussed previously.

In the experiments, the samples were cooled down to 1.8 K and the QPC transmittance was adjusted to $N=1$ 
for injector and detector. Figures~2, 3 and 4 show typical examples of measured non-local 
magnetoresistances. The observed structures are attributed to electron interference effects because of 
its marked temperature dependence (cf.\ Fig.~2). The interference patterns are stable in time and 
characteristic for a given sample and a given cooling cycle. Between different cooling cycles the 
interference pattern changes only slightly. This is in strong contrast with typical observations on 
electronic quantum interences, e.g.\ universal conductance fluctuations (UCF). UCF are related to 
electron scattering with single impurities, whose 'fingerprint' varies strongly from cooldown to 
cooldown, while, as discussed in the introduction, in high-mobility 2DEGs scattering is due to random 
potential fluctuations, which depend on the much more robust spatial charge correlations of donors.

\begin{figure}[T]
\begin{center}
\resizebox{6.5cm}{8cm}{\includegraphics{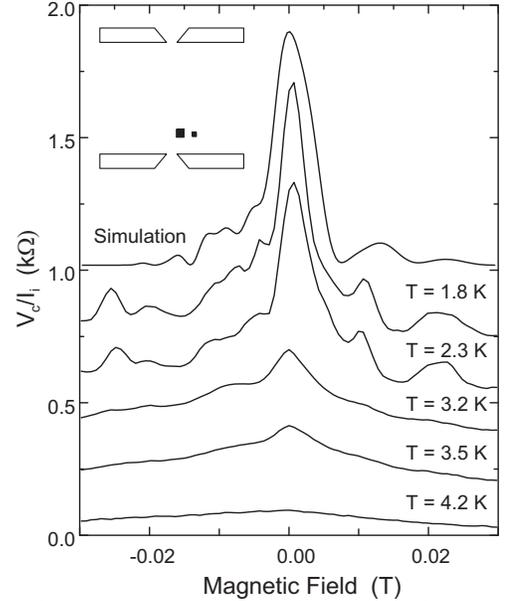}}
\begin{minipage}{8cm}
\caption{Non-local magnetoresistance at different temperature. A simulation of the experimental result 
(at 1.8 K) is obtained for a scatterer configuration indicated schematically in the top left corner: 
$L_{i,1}=0.6$ $\mu$m, $0.1 \le y_i \le 0.2$ $\mu$m and $L_{i,2}=0.6$, $\mu$m, $0.4 \le y_i \le 0.42$ 
$\mu$m.}
\label{temp}
\end{minipage}
\end{center}
\end{figure}

In order to substantiate our interpretation of the experimental observations it is now necessary to model 
the experimentally found diffraction patterns to gain information about size and location of scattering 
potentials. We use the simplest possible model, based on an extension of the method of Saito et 
al.\cite{saito}. The electron wavefunction at the exit of a QPC can be written as

\begin{equation}\label{wavein}
\Psi_0 (0,y)= \left(\frac{2}{W}\right)^{1/2} \, \cos\left(\frac{\pi y}{W}\right)\; \mbox{for} 
\, -W/2 \le y \le W/2 \; ,
\end{equation}

and zero elsewhere, if the QPC carries only one conducting mode (cf.\ Fig.~1b). $W$ denotes the point-
contact width at the exit. Using Green's theorem with Dirichlet's boundary conditions the wavefunction 
can be calculated for any point of the half-plane ($x>0$):

\begin{equation}
\Psi ({\bf r'})= \frac{i}{2m^\ast} \int_S dS \: {\bf n}({\bf r}) \cdot [ \Psi({\bf r}) (-i \hbar {\bf 
\nabla_r})  G^+({\bf r'},{\bf r})  ] \; .
\end{equation}

$G^+$, the Green's function in a weak magnetic field, which can be approximated by the Green's function 
at zero field, $G^0({\bf r',r})$, and a phase factor, $\theta ({\bf r',r })$:

\begin{eqnarray}
G^+ ({\bf r',r}) & \simeq & e^{i \theta ({\bf r',r})} \, G^0 ({\bf r',r}) \; ,\\
\theta ({\bf r',r })& = & - \frac{e}{\hbar} \int 
			{\bf A(R}(t)) \cdot \frac{{\bf \nabla_R} G^0({\bf R}(t),{\bf r})}{|{\bf \nabla_{R}} 
G^0({\bf R} (t),{\bf r})|}\, dt \; ,
\end{eqnarray} 

a line-integral along the gradient of $G^0({\bf r',r })$. For a detailed discription of this method we 
refer to Ref.~\cite{saito}. 
\begin{figure}
\begin{center}
\resizebox{6.5cm}{8cm}{\includegraphics{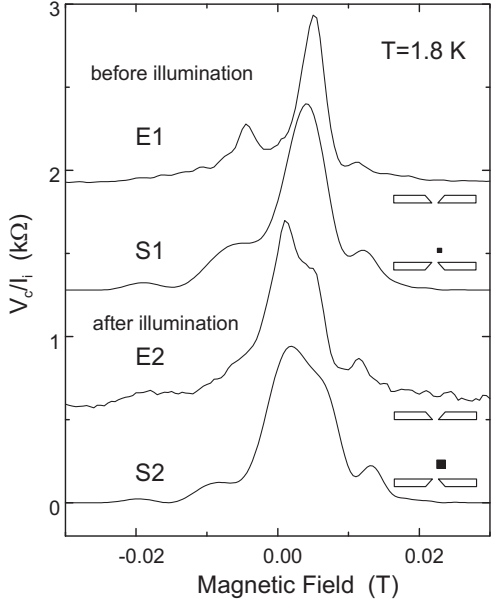}}
\begin{minipage}{8cm}
\caption{Experiment (E) and simulation (S) of the collimation signal before (1) and after illumination 
(2). The fit parameter for traces S1 and S2 are: $L_i=0.65$ $\mu$m, $0.08 \le y_i \le 0.16$ $\mu$m and 
$L_i=0.65$ $\mu$m, $0.08 \le y_i \le 0.22$ $\mu$m, respectively.}
\label{illumination}
\end{minipage}
\end{center}
\end{figure}
The wavefunction in the detector QPC, $\Psi_D (L,y)$, at a distance $x=L$ can be written analogous to 
Eq.\ \ref{wavein} (for one conducting mode). Thus, the transmission coefficient from the injector to the 
detector QPC can be calculated:

\begin{equation}\label{T}
T = \left| \int_{-W/2}^{W/2} \Psi_D^*(l,y')\Psi(l,y')dy'\right|^2 \; .
\end{equation}

In order to simulate impurities, an intermediate line is introduced between injector and detector QPC, 
($0 < x = L_i < L$). The wavefunction $\Psi_i$ is calculated at this line. The simplest model for the 
effect of a scattering object on the electronic wavefunction is just to set a part of the wavefunction 
$\Psi_i$ to zero, schematically shown in Fig.~1b. This modified wavefunction is then propagated further 
to calculate the detector wavefunction $\Psi_D$; Eq.~\ref{T} then gives the transmission probability. Of 
course, cutting off parts of a wavefunction is a very crude methode to simulate scattering. Neither 
diffusive back- or forward-scattering nor wavefunction matching at the boundaries are considered. 
However, a comparision with the experiment shows that this model yields a reasonable reproduction of the 
main features of the observed interference patterns (see Fig.~2). 

From fitting the theoretical curves to the experimental traces it is possible to deduce values for the 
distance $L_i$ and the width $W_i$ of scattering potential, yielding two scattering centers for the 
example displayed in Fig.~2: $L_{i,1}=0.6$ $\mu$m, $W_{i,1}=0.1$ $\mu$m ($0.1 \le y_i \le 0.2$ $\mu$m) 
and $L_{i,2}=0.6$ $\mu$m, $W_{i,2}=0.02$ $\mu$m ($0.4 \le y_i \le 0.42$ $\mu$m). The configuration of the 
scattering objects is schematically shown in the inset of Fig.~2. Already small variation 
\begin{figure}
\begin{center}
\resizebox{6.5cm}{8cm}{\includegraphics{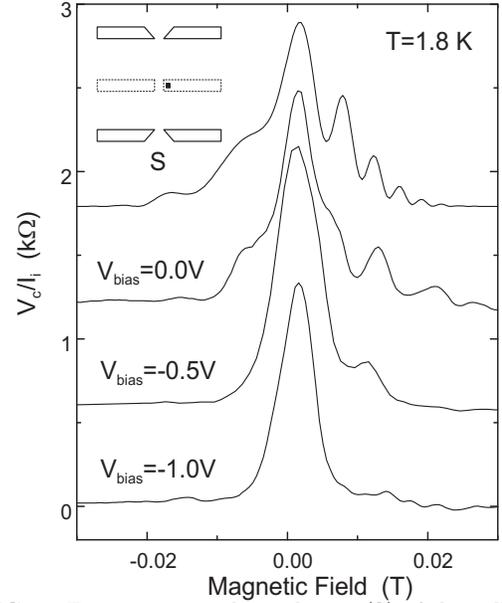}}
\begin{minipage}{8cm}
\caption{Experiment and simulation (S) of the collimation signal for the bias-cooling experiment. The 
applied voltages are indicated in the figure. The fit parameter for trace S is: $L_i=2.00$ $\mu$m and 
$0.18 \le y_i \le 0.23$ $\mu$m.}
\label{bias-cooling}
\end{minipage}
\end{center}
\end{figure}
of the size or location has a drastic effect on the observed interference patterns. From comparing the 
numerical results we estimate the uncertainty of the fitted values for $L_i$ and $W_i$ as less than 5\%.

Now experiments were done to prove that the diffracting objects are due to correlations in the 
distribution of donor states. Fig.~3, trace E1, shows the observed interference pattern for a sample 
cooled down in the dark. At 1.8 K the QPCs were defined and the collimation signal was measured. The 
simulation yielded a width of $W_{i,S1}=0.08$ $\mu$m in a distance $L_{i,S1}=0.65$ $\mu$m. Subsequently, 
the device was illuminated using a 100~$\mu$s light pulse of a red light-emitting diode ($\lambda 
=670$~nm) close to the sample. The resulting interference pattern differs significantly from the initial 
(Fig.~3, trace E2). Within the model for DX-center formation described above, this change can be 
attributed to a light induced transformation of donors in the DX$^-$-state into the d$^+$-state. At the 
same time the sample exhibits a slight increase in the carrier density, in good agreement with a DX$^-$ 
$\rightarrow$ d$^+$ conversion. From the simulation S2 an increase in the width of the scattering 
potential is found ($W_{i,S2}= 0.14$ $\mu$m). The observed change in the interference pattern of the 
electron beam is direct evidence for a reconfiguration of the scattering potential in the vicinity of the 
electron beam.

In Fig.~4 experimental curves are shown for a sample with an extra pair of gates between injector and 
detector. As indicated in Fig.~1a, these intermediate gates have a small gap of approximate 300 nm 
centered at the line connecting injector and detector. The collimation signal of this sample, cooled down 
without the intermediate gates defined, exhibits a distinct interference pattern. The simulation S 
reveals a scattering potential in the vicinity of the electron beam just underneath the intermediate gate 
($L_i = 2.00$ $\mu$m and $0.18 \le y_i \le 0.23$ $\mu$m) for $V_{\rm bias}=0$~V). As demonstrated in 
Ref.~\cite{buks} one may suppress the formation of DX-centers by depleting the 2DEG through the 
application of a negative bias voltage at high temperatures. Below $T = 130$~K this (uncorrelated) donor 
configuration will be stable against thermal activation and the bias voltage can be released. The 
interference patterns resulting from applying exactly this procedure to intermediate gates are shown in 
Fig.~4 for $V_{\rm bias}= 0.0$, $-0.5$ and $-1.0$~V. The initial interferences are suppressed with 
increasing negative bias voltage. This can be understood by considering that the applied bias voltage 
suppresses the formation of DX$^-$-Centers underneath the gates, leading to a smoothening of the 
potential landscape in these regions.

A common result of all simulations is that the size of the scattering potential is in the order of 50 to 
150~nm, comparable to those found in selfconsistent calculations for the size of potential fluctuations 
in a 2DEG due to a random distribution of correlated donors\cite{nixon,efros}. It is therefore very 
likely that the observed interference effects are related to donor complexes. Additional evidence for 
this conclusion is that modifying the distribution of charged donors by illumination or bias-cooling 
immediately effects the observed electron-beam interferences.

In conclusion, electron-beam experiments probe directly the existance of long range correlations between 
donors in GaAs-(Al,Ga)As heterostructure on a microscopic level. We used a numerical methode to simulate 
scattering potentials in the path of the electron beam in 2DEG layer. It was possible to deduce the 
position and size of actual scattering potentials by fitting experimentally obtained interference pattern 
of an electron-beam signal. The typical size of the scatterers ($50\dots 150$~nm) implies a collective 
effect of randomly distributed donors. This distribution could be changed by reducing the number of DX$^-
$-centers through illumination and bias-cooling techniques. The experiments show that a collimated 
electron beam is a sensitive tool in the investigation of local potential fluctuations in a 2DEG. It 
would be of interest to develop a more sophisticated theory of electron-beam scattering. Simulations 
using such a theory could in combination with e.g.\ density-dependent experiments on the interference 
structures be used to yield a detailed picture of the shape and size of the density fluctuations in a 
2DEG.
\vspace{-0.5cm}

\acknowledgments

The authers thank A.R.\ Long for discussions. Part of this work was supported by the DFG under MO 771/1-1 
and MO 771/3-1, and by the Volkswagen Stiftung.

\end{multicols}

\end{document}